\documentclass[12pt,epsfig]{article}
\usepackage{amsmath}
\usepackage{amsfonts}
\usepackage[dvips]{epsfig}
\psfigdriver{dvips}
\usepackage{mathrsfs} 
\newcommand{\ca}[1]{\mathscr{#1}}
\newcommand{\bo}[1]{\boldsymbol{#1}}
\newcommand{\fr}[1]{\mathfrak{#1}}

\begin{document}

%
\def\papertitlepage{\baselineskip 3.5ex \thispagestyle{empty}}
\def\preprinumber#1#2{\hfill \begin{minipage}{4.2cm}  #1
                 \par\noindent #2 \end{minipage}}
\renewcommand{\thefootnote}{\fnsymbol{footnote}}
\newcommand{\beq}{\begin{equation}}
\newcommand{\eeq}{\end{equation}}
\newcommand{\beqa}{\begin{eqnarray}}
\newcommand{\eeqa}{\end{eqnarray}}
\catcode`\@=11
\@addtoreset{equation}{section}
\def\theequation{\thesection.\arabic{equation}} 
\catcode`@=12
\relax
\newcommand{\ad}{\operatorname{ad}}
\newcommand{\mult}{\operatorname{mult}}

%
%
\papertitlepage
\setcounter{page}{0}
\preprinumber{}{hep-th/0512092}
\baselineskip 0.8cm
\vspace*{2.0cm}
\begin{center}
{\large\bf Five-dimensional Supergravity and Hyperbolic Kac-Moody Algebra 
$\bo{G_2^H}$}
\end{center}
\vskip 4ex
\baselineskip 1.0cm
\begin{center}
           {Shun'ya~ Mizoguchi${}^1$, Kenji Mohri${}^2$ 
and Yasuhiko Yamada${}^3$} 
\\
\vskip 1em
       ${}^1${\it High Energy Accelerator Research Organization (KEK)} \\
       \vskip -2ex {\it Tsukuba, Ibaraki 305-0801, Japan} \\
\vskip 1em
       ${}^2${\it Institute of Physics, University of Tsukuba} \\
       \vskip -2ex {\it Tsukuba, Ibaraki 305-8571, Japan}\\
\vskip 1em
      ${}^3$ {\it Department of Mathematics, Kobe University} \\
        \vskip -2ex {\it Rokko, Kobe 657-8501, Japan}
\end{center}
\vskip 5ex
%
\baselineskip=3.5ex
\begin{center} {\bf Abstract} \end{center}
Motivated by the recent analysis of the $E_{10}$ sigma model 
for the study of  M theory, we study a one-dimensional sigma model 
associated with the hyperbolic Kac-Moody algebra $G_2^H$ and its link 
to $D=5$, ${\cal N}=2$ pure supergravity,  which closely 
resembles  in many ways $D=11$ supergravity.
The bosonic equations of motion and the Bianchi identity for $D=5$ 
pure supergravity match the equations of the level $\ell\leq 3$ truncation of 
the $G_2^H$ sigma model up to higher level terms, just as 
they do for the $D=11$ case. We also compute low level root and outer 
multiplicities in the $A_3$ decomposition, and indeed find singlets   
at $\ell=4k$, $k=2,3,\ldots$ corresponding to the scaling of $ER^{k+1}$ 
terms, although the missing singlet at $\ell =4$ remains a puzzle.
\vskip 2ex
\vspace*{\fill}
\noindent
December 2005
\newpage
\renewcommand{\thefootnote}{\arabic{footnote}}
\setcounter{footnote}{0}
\setcounter{section}{0}
\baselineskip = 0.6cm
\pagestyle{plain}

\section{Introduction}
The recent study of gravity and supergravity solutions near a spacelike 
singularity has revealed that their oscillatory behavior 
is related in a not well understood way to consistency of superstring 
and M theories. A typical observation is that 
in the BKL limit \cite{BKL} pure gravity 
ceases to be chaotic in dimensions $D\geq 11$ \cite{Demaret}, 
but the oscillatory behavior 
is restored if gravity is coupled to a three form \cite{DHChaos}. 
The criterion of whether 
the given theory is chaotic or not can be summarized in a word : 
hyperbolicity \cite{DHJN}. 

In the $D=11$ supergravity case, the behavior of the logarithmic scale 
factors of metric is described as a billiard motion in the Weyl chamber of 
the hyperbolic Kac-Moody algebra $E_{10}$ \cite{DNE10BE10}. 
A systematic analysis of 
this `cosmological billiard' was carried out in \cite{CosmologicalBilliard},
in which it was shown that the billiard dynamics is asymptotically equivalent 
to a one-dimensional sigma model associated with a corresponding 
hyperbolic Kac-Moody group. 
(See \cite{Chitre}-\cite{Ivashchuk:1999rm} for the pioneering works on the billiard approach.)
Moreover, agreement was found between 
equations of motion of $D=11$ supergravity and those of the $E_{10}$ 
sigma model up to height $\leq 29$ in the framework of the $A_9$ `level' 
decomposition of $E_{10}$ \cite{DHNsmalltension,NF,DN}
(See also \cite{Axel} for $A_d$ decomposition of the very extended Kac-Moody 
algebras.), 
which has led to the conjecture 
that even information on higher order corrections of M theory is encoded 
in the infinite towers of roots of $E_{10}$ 
(See \cite{Nicolaigr-qc} for the relevance of $E_{10}$ in M theory. 
See also \cite{BGH} for a different M-theory interpretation of the 
imaginary roots.). Similar analyses were also done 
in massive IIA \cite{KNmassiveIIA} and IIB supergravities \cite{KNIIB}.

More recently, further evidence supporting this conjecture was given in 
the $E_{10}/A_9$ decomposition analysis \cite{DNhigherorder}, 
in which a series of $A_9$ singlets, 
whose existence was suggested by the scaling behavior of higher derivative 
corrections of the form $ER^N$, 
were indeed found at levels $\ell=10k$, $k=1,2,\ldots$.
This predicted that the higher order corrections are allowed only for 
$N=4,7,\ldots$, which was recently confirmed to be consistent with 
the string duality \cite{GV}. 

As was emphasized in \cite{DNhigherorder}, the correspondence between the 
wall forms associated with the higher curvature corrections 
and singlet representations of the relevant subalgebra appears 
to be a special property of M theory and $E_{10}$, 
which is not shared by, for instance, pure Einstein gravity and $AE_d$ 
\cite{Nicolaigr-qc}. Therefore it will be interesting
to explore if there is any other such supergravity model which also possesses 
singlets in the decomposition at the locations expected from the scaling 
behavior of the higher curvature corrections.  
In this paper we use $D=5$ pure supergravity \cite{D=5} 
as our example to investigate.

It is known that $D=11$ supergravity \cite{D=11} and $D=5$ pure supergravity  
are very similar \cite{MO} in many ways. In particular, 
the dimensional reduction to three dimensions can be done in a very similar 
manner in both theories to obtain 
$E_{8(+8)}/SO(16)$ \cite{MS,MizoE10} and $G_{2(+2)}/SO(4)$ \cite{MO} 
nonlinear sigma models, respectively.
Moreover, it was shown \cite{DdBHS} that $D=5$ pure supergravity 
is also one of the special class of theories that exhibits chaotic behavior 
in the BKL limit, in which the billiard is the Weyl chamber of $G_2^H$, 
the canonical hyperbolic extension of $G_2$. 
The similarity of $E_8$ and $G_2$ was noted in \cite{MizoGermar}. 
Therefore we consider, in place of $E_{10}$, a one-dimensional sigma model 
associated with $G_2^H$, and study its relation to $D=5$ pure supergravity.  
We will see, again, that there is a strong parallelism. 
We will show that  
the bosonic equations of motion and the Bianchi identity for $D=5$ 
pure supergravity match the equations of the level $\ell\leq 3$ truncation of 
the $G_2^H$ sigma model up to a few higher level terms, just as 
they do \cite{DN} for the $D=11$ case. 
We also compute low level root and outer 
multiplicities in the $A_3$ decomposition up to height$\leq 40$ 
and $\leq60$, respectively, 
and find singlets  at $\ell=4k$, $k=2,3,\ldots$, which precisely corresponds 
to the scaling of $ER^{k+1}$ terms in $D=5$ pure supergravity. 
However, it turns out that there are no singlets at $\ell=4$, which is puzzling
because there is a corresponding on-shell 1-loop divergence;
this will be discussed in Conclusions.

\section{The hyperbolic Kac-Moody algebra $\bo{G_2^H}$}
The Kac-Moody algebra $G_2^H$ is defined to be generated by multiple 
commutators of the Chevalley generators $\{e_i,f_i,h_i\}$ $(i=0,1,2,3)$ 
with the relations
\beqa
&&{[}h_i,~e_j{]}=A_{ij} e_j, ~~~{[}h_i,~f_j{]}=-A_{ij} f_j, 
~~~{[}e_i,~f_j{]}=\delta_{ij} h_i,
\nonumber\\
&&~~~~~(\ad e_i)^{1-A_{ij}}(e_j)=0, ~~~(\ad f_i)^{1-A_{ij}}(f_j)=0,
\eeqa
where 
$A_{ij}$ is the Cartan matrix
\beqa
A_{ij}&=&\frac{2(\alpha_i|\alpha_j)}{(\alpha_i|\alpha_i)}~=~\left(
\begin{array}{rrrr}
2&-3&0&0\\
-1&2&-1&0\\
0&-1&2&-1\\
0&0&-1&2
\end{array}
\right).
\label{Cartan}
\eeqa
$\alpha_i$ ($i=0,\ldots,3$) are the simple roots.
\vspace{0.5cm}

\begin{figure}[h]
\begin{center}
\epsfbox{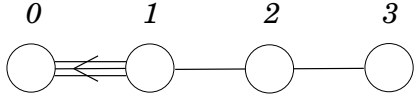}
\end{center}
\caption{Dynkin diagram of $G_2^H$}
\end{figure}
%
$G_2^H$ has a regular subalgebra $A_3$ whose simple roots consist of  
$\{\alpha_1,\alpha_2,\alpha_3\}$. 
We decompose the whole set of $G_2^H$ 
roots into irreducible orbits of  this $A_3$ action.

Any root of $G_2^H$ can be written as
\beqa
\alpha &=& \ell\alpha_0 +\sum_{j=1}^3 m^j \alpha_j
\label{alpha}
\eeqa
with all non-negative or all non-positive integers $\ell$ and $m^j$. 
The coefficient 
$\ell$ is called the {\it  level} \cite{DHNsmalltension} of $\alpha$. 
By definition, the $A_3$ 
action does not change $\ell$; the whole set of roots at each level $\ell$ 
are decomposed into a direct sum of $A_3$ representations, which are 
specified by their Dynkin labels 
$p_k(\varLambda)\equiv(\alpha_k|\varLambda)\quad (k=1,2,3)$
of some highest weight $\varLambda$.

We can proceed exactly in the same way as the $E_{10}/A_9$ or
$AE_3/A_2$ decomposition.
For example, the Chevalley generator $f_0$ is a root vector with root 
$-\alpha_0$, which is the highest weight vector of $A_3$ with Dynkin label 
$(p_1,p_2,p_3)=(1,0,0)$; all the level $\ell=-1$ roots corresponds 
in one to one to its weights,
and hence are components of an $A_3$ vector. 
In the next section we will see that 
it is identified as the (spatial part of) $U(1)$ gauge field in $D=5$ simple 
supergravity.

One can also derive constraints in order 
for a negative root $-\alpha$ (\ref{alpha}) to be a highest weights of some 
representation of the $A_3$ subalgebra, similarly to \cite{DHNsmalltension}. 
The result is, 
for $\varLambda= -\alpha$,
\begin{align}
p_1&=\ell-2m^1+m^2~\geq0,\nonumber\\
p_2&=m^1-2 m^2 +m^3~\geq0,\label{constraint1}\\
p_3&=m^2-2 m^3 ~\geq0,\nonumber
\end{align}
and
\begin{equation}
|\alpha|^2=-\frac1{12}\ell^2+\frac14\left(
3p_1^2+4p_2^2+3p_3^2+4p_1p_2 +4p_2p_3 + 2p_1p_3
\right)~\leq 2.
\label{constraint2}
\end{equation}
Using these constraints one can easily show that there is a 
unique dominant weight\\
 $(p_1,p_2,p_3)=(0,1,0)$ at $\ell=-2$, which 
corresponds to a $A_3$ rank-2 antisymmetric tensor. 
We will see that it corresponds to the electro-magnetic dual 
of the gauge field.

At $\ell=-3$, there are two solutions $(p_1,p_2,p_3)=(1,1,0)$ and 
$(0,0,1)$ to the constraints (\ref{constraint1}),(\ref{constraint2}).
As we will see, however,  the latter is not a highest weight,
and there is again a unique representation at $\ell=-3$.
It carries three mixed symmetric and anti-symmetric indices and 
is identified as a dual graviton in Section 3.

\newpage
\renewcommand{\arraystretch}{1.2}
\begin{center}
Table 1.  The three lowest level representations of the $E_{10}/A_9$
and $G_2^H/A_3$ decompositions.\\
\vskip 5mm
\begin{tabular}{|r|c|c|}
\hline
$\ell$ & $A_9$ Dynkin label & $A_3$ Dynkin label\\

\hline

$-1$&$(0,0,1,0,0,0,0,0,0)$&$(1,0,0)$\\
$-2$&$(0,0,0,0,0,1,0,0,0)$&$(0,1,0)$\\
$-3$&$(1,0,0,0,0,0,0,1,0)$&$(1,1,0)$\\
\hline
\end{tabular}

\end{center}

These three lowest representations 
in our $G_2^H/A_3$ decomposition show a striking resemblance to 
those in the $E_{10}/A_9$ decomposition \cite{DHNsmalltension} 
in their index structures,
being a reflection of the similarity between $D=11$ and $D=5$ supergravities.

Since the constraints can only provide necessary conditions for 
the highest weights,
in order to further compute the decomposition, we employ Peterson's 
recursive formula for root multiplicities \cite{Kac}
\beqa
(\beta|\beta-2\rho)c_\beta&=&
\sum_{\begin{subarray}{l}
\beta',\beta''\in Q_+\\
\beta'+\beta''=\beta
\end{subarray}}
(\beta'|\beta'')
c_{\beta'}c_{\beta''},
\eeqa
where $
c_{\beta}=\sum_{n\geq 1} n^{-1}\mult(\beta/n)$ and 
$Q_+=\sum_{i=0}^3 {\mathbb{Z}}_{+} \alpha_i$. We have used 
{\tt Mathematica} to obtain root multiplicities of $G_2^H$ with low 
heights. 
Results of a sample computation are shown in Table 2, in which
multiplicities of all the {\it positive} roots 
$\alpha=\ell\alpha_0+m^1\alpha_1+m^2\alpha_2+m^3\alpha_3$ 
that are {\it dominant} as 
an $A_3$ weight (that is, all the Dynkin labels $p_i$ are non-negative) 
are listed 
up to $\mbox{height}\leq40$.\footnote{We could equally compute multiplicities 
of {\it negative} roots.
However, although the number of $A_3$-dominant $G_2^H$ roots at level $\ell$ is 
of course the same as that at level $-\ell$, the height of the former is 
generally larger than the latter, so we would have to display more roots 
for the same given
maximum height $(=40)$.}

At $\ell=0$, we can see the highest weight of the adjoint representation 
of $A_3$ as expected. For $\ell=1,2$ there is a unique representation  
at each level. These dominant weights are minus the lowest weights 
in the corresponding representations at $\ell=-1,-2$ in Table 1. 
At $\ell=3$, we have $(0,1,1)$ and $(1,0,0)$ with 
root multiplicities 1 and 2, respectively. The former representation 
contains $(1,0,0)$ as one of the weights, with weight multiplicity being 
precisely two. Therefore 
the {\it outer multiplicity} (that is, how many times the representation 
occurs) of $(1,0,0)$ is zero.\footnote{The level decomposition for the `very-extended'
version of $G_2$ has been worked out in \cite{Axel}. At low levels there is no difference between 
$G_2^H$ and their result. In particular one can find the supergravity fields as well as the same 
outer multiplicities of the $A_3$ singlets at $\ell=4$ and $8$. We thank Axel Kleinschmidt 
for pointing this out.
}


\newpage
\renewcommand{\arraystretch}{1.1}
\begin{center}
Table 2. ~A sample computation of root multiplicities of $G_2^H$ 
at low height.\\
\vskip 5mm
\mbox{\small 
\begin{tabular}{|l|r|r|l|}
\hline
$(\ell,m^1,m^2,m^3)$ & $3\alpha^2$ & mult & $(p_1,p_2,p_3)$\\
\hline

$(0,1,1,1)$&$6
$&$1$&$(
1,0,1)$\\

$(1,1,1,1)$&$2
$&$1$&$(
0,0,1)$\\

$(2,2,2,1)$&$2
$&$1$&$(
0,1,0)$\\

$(3,3,2,1)$&$0
$&$2$&$(
1,0,0)$\\

$(4,3,2,1)$&$-4
$&$3$&$
(0,0,0)$\\

$(3,3,3,2)$&$6
$&$1$&$(
0,1,1)$\\

$(4,4,3,2)$&$2
$&$1$&$(
1,0,1)$\\

$(5,4,3,2)$&$-4
$&$3$&$(
0,0,1)$\\

$(5,5,4,2)$&$2
$&$1$&$(
1,1,0)$\\

$(6,5,4,2)$&$-6
$&$6$&$(
0,1,0)$\\

$(6,6,4,2)$&$0
$&$2$&$(
2,0,0)$\\

$(6,5,4,3)$&$0
$&$2$&$(
0,0,2)$\\

$(7,6,4,2)$&$-10$&$9$&$
(1,0,0)$\\

$(8,6,4,2)$&$-16$&$21$&$
(0,0,0)$\\

$(6,6,5,3)$&$6
$&$1$&$(
1,1,1)$\\

$(7,6,5,3)$&$-4
$&$3$&$(
0,1,1)$\\

$(7,7,5,3)$&$2
$&$1$&$(
2,0,1)$\\

$(8,7,5,3)$&$-10$&$9$&$
(1,0,1)$\\

$(9,7,5,3)$&$-18
$&$32$&$(
0,0,1)$\\

$(8,7,6,3)$&$-4
$&$3$&$(
0,2,0)$\\

$(8,8,6,3)$&$2
$&$1$&$(
2,1,0)$\\

$(8,7,6,4)$&$2
$&$1$&$(
0,1,2)$\\

$(9,8,6,3)$&$-12
$&$14$&$(
1,1,0)$\\

$(10,8,6,3)$&$-22$&$48$&$
(0,1,0)$\\

$(9,9,6,3)$&$0
$&$2$&$(
3,0,0)$\\

$(9,8,6,4)$&$-6
$&$6$&$(
1,0,2)$\\

$(10,9,6,3)$&$-16$&$21$&$
(2,0,0)$\\

$(10,8,6,4)$&$-16$&$21$&$
(0,0,2)$\\

$(9,8,7,4)$&$0
$&$2$&$(
0,2,1)$\\

$(11,9,6,3)$&$-28$&$99$&$
(1,0,0)$\\

$(9,9,7,4)$&$6
$&$1$&$(
2,1,1)$\\

$(12,9,6,3)$&$-36
$&$258$&$
(0,0,0)$\\

$(10,9,7,4)$&$-10$&$9$&$
(1,1,1)$\\

$(11,9,7,4)$&$-22$&$48$&$
(0,1,1)$\\

\hline
\end{tabular}
%
%
%
\begin{tabular}{|l|r|r|l|}
\hline
$(\ell,m^1,m^2,m^3)$ & $3\alpha^2$ & mult & $(p_1,p_2,p_3)$\\
\hline

$(10,10,7,4)$&$2$&$1$&$
(3,0,1)$\\

$(10,9,8,4)$&$2
$&$1$&$(
0,3,0)$\\

$(10,9,7,5)$&$2
$&$1$&$(
1,0,3)$\\

$(11,10,7,4)$&$-16$&$21$&$
(2,0,1)$\\

$(11,9,7,5)$&$-10$&$9$&$
(0,0,3)$\\

$(12,10,7,4)$&$-30
$&$135$&$
(1,0,1)$\\

$(11,10,8,4)$&$-10$&$9$&$
(1,2,0)$\\

$(13,10,7,4)$&$-40$&$378$&$
(0,0,1)$\\

$(12,10,8,4)$&$-24
$&$66$&$(
0,2,0)$\\

$(11,11,8,4)$&$2$&$1$&$
(3,1,0)$\\

$(11,10,8,5)$&$-4$&$3$&$
(1,1,2)$\\
$(12,11,8,4)$&$-18
$&$32$&$(
2,1,0)$\\

$(12,10,8,5)$&$-18
$&$32$&$(
0,1,2)$\\

$(13,11,8,4)$&$-34$&$199$&$
(1,1,0)$\\

$(12,12,8,4)$&$0
$&$2$&$(
4,0,0)$\\

$(12,11,8,5)$&$-12
$&$14$&$(
2,0,2)$\\

$(12,10,8,6)$&$0
$&$2$&$(
0,0,4)$\\

$(14,11,8,4)$&$-46$&$702$&$
(0,1,0)$\\

$(13,12,8,4)$&$-22$&$48$&$
(3,0,0)$\\

$(13,11,8,5)$&$-28$&$99$&$
(1,0,2)$\\

$(12,11,9,5)$&$-6
$&$6$&$(
1,2,1)$\\

$(14,12,8,4)$&$-40$&$378$&$
(2,0,0)$\\

$(14,11,8,5)$&$-40$&$378$&$
(0,0,2)$\\

$(13,11,9,5)$&$-22$&$48$&$
(0,2,1)$\\

$(12,12,9,5)$&$6
$&$1$&$(
3,1,1)$\\

$(12,11,9,6)$&$6
$&$1$&$(
1,1,3)$\\

$(15,12,8,4)$&$-54
$&$1559$&$
(1,0,0)$\\

$(13,12,9,5)$&$-16$&$21$&$
(2,1,1)$\\

$(13,11,9,6)$&$-10$&$9$&$
(0,1,3)$\\

$(16,12,8,4)$&$-64$&$3786$&$
(0,0,0)$\\

$(14,12,9,5)$&$-34$&$199$&$
(1,1,1)$\\

$(13,13,9,5)$&$2$&$1$&$
(4,0,1)$\\

$(13,12,10,5)$&$-4$&$3$&$
(1,3,0)$\\

$(13,12,9,6)$&$-4$&$3$&$
(2,0,3)$\\
\hline
\end{tabular}
}
\end{center}

%
{}From this table we observe the following two further similarities to 
the $A_9$ decomposition of $E_{10}$. First, $G_2^H$ has three 
towers of roots with $A_3$ Dynkin labels $(n,0,1)$, $(n,1,0)$ and $(n,1,1)$
at levels $\ell=3n+1$, $3n+2$ and $3n+3$, respectively, with root multiplicity 
one. (Their outer multiplicities are also one.)
 They are the $G_2^H$ analogue of the three series of $E_{10}$ 
roots with $A_9$ labels $(0,0,1,0,0,0,0,0,n)$, $(0,0,0,0,0,1,0,0,n)$ and 
$(1,0,0,0,0,0,0,1,n)$.\footnote{When compared with our $G_2^H$ roots, 
the ordering of Dynkin 
labels should be reversed since we are decomposing {\it positive} roots here.}
Thus we may say that there is also `enough room' \cite{DHNsmalltension}
in $G_2^H$ 
roots for the spatial gradients of $D=5$ supergravity fields identified at 
$\ell\leq 3$, just in the same manner as $E_{10}$ contains the spatial 
gradients of $D=11$ supergravity.

The second interesting observation is that there are a series of 
$A_3$ weights $(0,0,0)$  at levels $\ell=4k$, $k=1,2,\ldots$ with 
root coefficients $(\ell,m^1,m^2,m^3)=(4k,3k,2k,k)$. 
(In fact, the first one ($k=1$) turns out 
to have outer multiplicity zero.)
It was found in \cite{DNhigherorder} that $E_{10}$ has a series of 
singlets in the $A_9$ decomposition, and such roots are proportional 
to the `wall form' of higher order curvature corrections 
of $D=11$ supergravity.  
In section 4 we will discuss the relevance of these 
$A_3$ singlets to higher order corrections to $D=5$ supergravity.

\section{$\boldsymbol{G_2^H}$ sigma model and $\bo{D\!=\!5}$ supergravity}
\subsection{$\bo{G_2^H}$ generators for $\boldsymbol{\ell\leq 3}$}
In ref.\cite{DN} the comparison was made between the equations of motion of 
the $E_{10}/K(E_{10})$ sigma model and those of $D=11$ supergravity, 
in which the decomposition of $E_{10}$ under $A_9$ representations was 
used to show their matching up to `level' $\ell\leq 3$. 
In this section we will do 
a similar analysis for a $G_2^H/K(G_2^H)$ sigma model and $D=5$ supergravity
to find, again, a very similar result.

We write,  as in the case of $A_9$ in 
$E_{10}$, the generators of the $A_3$ subalgebra as
\beqa
{[}K^a_{~b},~K^c_{~d}{]}&=&\delta^c_b K^a_{~d}-\delta^a_d K^c_{~b}
\eeqa 
with, in our case, $a,b,\ldots=0,\ldots,3$. 
We also take two conjugate 
$A_3=sl(4)$ vectors $E^a$, $F_a$ transforming
\beqa
{[}K^a_{~b},~E^c{]}=\delta^c_b E^a,~~~
{[}K^a_{~b},~F_c{]}=-\delta^a_c F_b
\eeqa
($a=0,\ldots,3$). The relation between $F_a$ and $E^b$ is
\beqa
{[}F_a,~E^b{]}&=&-3 K^b_{~a}+\delta^b_a K,
\eeqa
where $K=K^1_{~1}+\cdots+K^4_{~4}$.
They are identified as the elements of $G_2^H$ which belong to 
the root spaces with $\ell=\pm1$. 
One can then take the Chevalley generators as 
\begin{alignat}{3}
&e_0=F_1, & \quad
&f_0=E^1, & \quad
&h_0=-3 {K^1}_{1}+K,\nonumber\\
&e_i={K^i}_{i+1}, & \quad
&f_i={K^{i+1}}_{i}, & \quad
&h_i={K^{i}}_{i}-{K^{i+1}}_{i+1} 
\end{alignat}
$(i=1,2,3)$. Note that no summation is taken over 
the repeated indices in the definition of $h_i$. 
We further define the $\ell=\pm2$ and $\pm3$ generators as
\begin{align}
E^{ab}&\equiv {[}E^a,~E^b{]},\nonumber\\
F_{ab}&\equiv -{[}F_a,~F_b{]},\nonumber\\
E^{b_0|b_1b_2}&\equiv {[}E^{b_0},~E^{b_1b_2}{]},\nonumber\\
F_{a_0|a_1a_2}&\equiv -{[}F_{a_0},~F_{a_1a_2}{]},
\end{align}
then
\beqa
{[}F_a,~E^{b_1b_2}{]}&=&8\delta_a^{[b_1}E^{b_2]},\nonumber\\
{[}F_{a_1a_2},~E^{b_1b_2}{]}&=&-16(3\delta^{[b_1}_{[a_1}K^{b_2]}_{~~a_2]}
-\delta^{[b_1}_{[a_1}\delta^{b_2]}_{a_2]}K),\nonumber\\
{[}F_a,~E^{b_0|b_1b_2}{]}&=&2(\delta_a^{b_0}E^{b_1b_2}-\delta_a^{[b_1}E^{b_2]b_0}),
\nonumber\\
{[}F_{a_1a_2},~E^{b_0|b_1b_2}{]}&=&
16(\delta_{[a_1}^{b_0}\delta_{a_2]}^{[b_1}E^{b_2]}
-\delta_{[a_1}^{[b_1}\delta_{a_2]}^{b_2]}E^{b_0}
),
\nonumber\\
{[}F_{c|a_1a_2},~E^{d|b_1b_2}{]}&=&
48\left[3\big(\delta^{[d}_c\delta^{b_1}_{[a_1}K^{b_2]}_{~~a_2]}
-\delta^{d}_c\delta^{[b_1}_{[a_1}K^{b_2]}_{~~a_2]}
-\delta^{[b_1}_{[a_1}\delta^{b_2}_{a_2]}K^{d]}_{~~c}
+\delta^{d}_{[a_1}\delta^{[b_1}_{a_2]}K^{b_2]}_{~~c}
\big)
\right.
\nonumber\\
&&\left.+\big(\delta_c^d\delta_{[a_1}^{[b_1}\delta_{a_2]}^{b_2]}
-\delta_c^{[b_1}\delta_{[a_1}^{b_2]}\delta_{a_2]}^{d}
\big)K\right].
\eeqa
The last equation can be conveniently written as 
\begin{equation}
{[}F_{c|a_1a_2},~X_{d|b_1b_2}E^{d|b_1b_2}{]}
=48
\left(3X_{c|b[a_1 } {K^b}_{a_2]}
+X_{c|a_1a_2}K
+3X_{[a_1|a_2]b}K^b_{~c}
-X_{[a_1|a_2]c}K
\right)
\end{equation}
for $X_{a|bc}$ satisfying  $X_{[a|bc]}=0$.

\subsection{$\bo{G_2^H}$ sigma model equations of motion}
Let $\theta$ be the Chevalley involution and 
define the transpose operation $T$ 
as $T=-\theta$. Let $\ca{V}(t)$ be a formal exponentiation of an $t$ dependent 
element of $G_2^H$. Let 
\beqa
\ca{V}^{-1}\partial_t \ca{V}&=&\ca{Q}_t + \ca{P}_t
\eeqa
with $\ca{Q}_t^T=-\ca{Q}_t$, $\ca{P}_t^T=+\ca{P}_t$.
Using the invariant bilinear form $\langle~ \cdot~|~\cdot~\rangle$ 
of the Kac-Moody algebra, 
we define the coset $G_2^H/K(G_2^H)$ sigma model Lagrangian as
\beqa
{\cal L}&=&\frac12 n^{-1}\langle\ca{P}_t |\ca{P}_t\rangle
\eeqa
with the lapse parameter $n^{-1}$,  where $G_2^H$ here is the corresponding 
Kac-Moody {\it group}, and 
$K(G_2^H)$ is the (formal) maximal compact subgroup 
whose Lie algebra is spanned by the `antisymmetric'  elements 
with respect to the 
afore-defined transposition $T$. 
The equation of motion derived from this Lagrangian is 
\beqa
n\partial_t(n^{-1}\ca{P}_t)+{[}\ca{Q}_t,~\ca{P}_t{]}&=&0,
\eeqa
where
\beqa
\ca{Q}_t &=& Q^{(0)}_t * L +\frac1{2} P^{(1)}_t *(E^{(1)}-F^{(1)}) +\cdots ,
\nonumber\\
\ca{P}_t &=& P^{(0)}_t * S +\frac1{2} P^{(1)}_t *(E^{(1)}+F^{(1)}) +\cdots ,
\eeqa
with $L_{ab}\equiv\frac12(K^a_{~b}-K^b_{~a})$, 
$S_{ab}\equiv\frac12(K^a_{~b}+K^b_{~a})$, 
and omitting the subscript $t$
\beqa 
Q^{(0)}* L &=&Q^{(0)}_{ab} L_{ab},\nonumber\\
P^{(0)} * S &=&P^{(0)}_{ab} S_{ab},\nonumber\\
P^{(1)} * E^{(1)} &=&P^{(1)}_{a} E^a,\nonumber\\
P^{(2)} * E^{(2)} &=&\frac1{2}P^{(2)}_{ab} E^{ab},\nonumber\\
P^{(3)} * E^{(3)} &=&\frac1{3!}P^{(3)}_{a|bc} E^{a|bc}.
\eeqa
$\ca{D}^{(0)}$ is defined to act on an $A_3$ vector $V_a$ as
\beqa
\ca{D}^{(0)}V_a &\equiv & \partial V_a+(Q^{(0)}_{ab}-P^{(0)}_{ab})V_b, 
\eeqa
and extends to tensors in the same way as an ordinary covariant derivative. 
There is no distinction between the upper and lower indices. Then
we have

\noindent
\underline{\it $\ell=0$ equation}
\beqa
n\ca{D}^{(0)}(n^{-1} P^{(0)}_{ab})&=&
-\frac32 \Big(P^{(1)}_aP^{(1)}_b -\frac13 \delta_{ab} P^{(1)}_cP^{(1)}_c\Big)
- 6 \Big(P^{(2)}_{ca}P^{(2)}_{cb} 
-\frac13 \delta_{ab} P^{(2)}_{cd}P^{(2)}_{cd}\Big)
\nonumber\\
&&-2 P^{(3)}_{c|da}P^{(3)}_{c|db} -P^{(3)}_{a|cd}P^{(3)}_{b|cd}
+ \delta_{ab} P^{(3)}_{c|de}P^{(3)}_{c|de} .
\eeqa

\noindent
\underline{\it $\ell=1$ equation}
\beqa
n\ca{D}^{(0)}(n^{-1} P^{(1)}_{a})&=&
-4 P^{(2)}_{ab}P^{(1)}_b +4 P^{(3)}_{b|ca}P^{(2)}_{bc}.
\eeqa

\noindent
\underline{\it $\ell=2$ equation}
\beqa
n\ca{D}^{(0)}(n^{-1} P^{(2)}_{ab})&=&
P^{(3)}_{c|ab}P^{(1)}_{c}.
\eeqa

And also the trivial $\ell=3$ equation 
$n\ca{D}^{(0)}(n^{-1} P^{(3)}_{a|bc})=0$.

\subsection{Comparison with $\bo{D\!=\!5}$ supergravity}
The bosonic Lagrangian for $D=5$, ${\cal N}=2$ pure gravity is
\beqa
{\cal L}_{\text{SUGRA}}=E\Big(
R-\frac34 F_{MN} F^{MN}
\Big)
-\frac14 \epsilon^{MNPQR}F_{MN}F_{PQ}A_R.
\eeqa
We have taken an unconventional normalization for the vector kinetic term 
so that the equations of motion are simplified. The relevant equations are 

\noindent
\underline{\it Einstein's equation}
\beqa
R_{AB}&=&\frac32 F_A^{~C} F_{BC}-\frac14 \eta_{AB} F^2.
\eeqa

\noindent
\underline{\it Maxwell's equation}
\beqa
D_AF^{AB}&=&\frac12 \epsilon^{BCDEF} F_{CD} F_{EF}.
\eeqa

\noindent
\underline{\it Bianchi identity}
\beqa
D_{[A}F_{BC]}&=&0.
\eeqa

We take the pseudo-Gaussian gauge 
\beqa
E_M^{~A}&=&\left(\begin{array}{cc}
N&0\\
0&e_m^{~a}
\end{array}\right)
\eeqa
($\det e_m^{~a}\equiv e$, $n\equiv N e^{-1}$)
and make the following 
Identification :
\beqa
P^{(0)}_{ab}&=&N\omega_{ab0},\nonumber\\
Q^{(0)}_{ab}&=&N\omega_{0ab},\nonumber\\
P^{(1)}_{a}&=&NF_{0a},\nonumber\\
P^{(2)}_{ab}&=&-\frac14 N\epsilon_{abcd} F_{cd},\nonumber\\
P^{(3)}_{a|bc}&=&-\frac14 N\epsilon_{bcde} \widetilde{\Omega}_{dea}. \nonumber\\
\eeqa
The coefficients of anholonomy 
$\Omega_{AB}^{~~~C}\equiv 2E_A^{~M}E_B^{~N}\partial_{[M}E_{N]}^{~~C}$ are 
decomposed as
\beqa
\Omega_{abc}\equiv \widetilde\Omega_{abc}+\frac23\Omega_{[a}\delta_{b]c},~~~
\Omega_a\equiv \Omega_{abc}\delta^{bc}
\eeqa
and set $\Omega_a=0$ as was done in \cite{DN}. Then it can be shown that
\begin{itemize}
\item{Einstein's equation coincides with the $\ell=0$ equation up to terms \\
$\displaystyle{-\frac12(
\partial_c\widetilde{\Omega}_{cab}+\partial_c\widetilde{\Omega}_{cba}
+\widetilde{\Omega}_{acd}\widetilde{\Omega}_{bdc})}$ (extra in $R_{ab}$).
}
\item{Maxwell's equation coincides with the $\ell=1$ equation up to a term \\
$-N^{-1}\partial_b(NF_{ba})$ (extra in $DF$).}
\item{Bianchi identity coincides with the $\ell=2$ equation up to a term \\
$2N^{-1}\partial_{[a}(NF_{b]0})$ (extra in $DF$).}
\end{itemize}
All the terms that do not match are similar in their structure to those in 
the $D=11$ case, and can be regarded as higher level contributions 
by the same scaling argument as in \cite{DN}.

\section{Singlet representations and higher order corrections}
\subsection{Wall forms for higher order corrections}
In the BKL limit the cosmological billiard of $D=5$ supergravity 
coincides with a scaling limit of the $G_2^H$ sigma model.
This follows from the general theorem for the cosmological billiard 
\cite{CosmologicalBilliard}
proven using the Iwasawa decomposition\footnote{
In this section, summation is not taken over the repeated indices 
but written explicitly.
}
\beqa
e^{a}_{~i}&=&\exp(-\beta^a )\cdot N^{a}_{~i} 
\eeqa
($i=0,\ldots,3$).
In the limit the off-diagonal degrees of freedom tend to freeze 
asymptotically, leaving a one-dimensional sigma model
with metric 
\beqa
\sum_{a,b=0}^3 G_{ab}\beta^a\beta^b
&=&\sum_{a=0}^3(\beta^a)^2 -\left(
\sum_{a=0}^3 \beta^a
\right)^2,
\eeqa
which is coupled to sharp `wall' potentials forming a billiard.
The inverse metric $G^{ab}$ is transformed by the `wall form matrix' 
\cite{DdBHS} 
\beqa
U_{ia}=\left(
\begin{array}{rrrc}
1&1&1&~~0~\\
-1&1&0&~~0~\\
0&-1&1&~~0~\\
0&0&-1&~~1~
\end{array}
\right)
\eeqa
to the matrix of bilinear pairing of $G_2^H$ roots 
\beqa
\sum_{a,b=0}^3 U_{ia}U_{jb}G^{ab}
&=&(\alpha_i|\alpha_j),
\eeqa
where the lhs is given by (\ref{Cartan}) with a normalization
$(\alpha_1|\alpha_1)=2$. In the BKL limit, the leading behavior 
of the supergravity metric is governed by a billiard motion in a 
chamber enclosed by sharp exponential potentials of the 
wall forms given by $U_{ia}$ \cite{DdBHS}.

On the other hand, in the analysis of the $E_{10}$ sigma model, 
the higher order curvature corrections $ER^4,ER^7,\ldots$ are 
identified to correspond to some negative roots of $E_{10}$ which 
are singlet under the $A_9$ decomposition. The identification was 
made by estimating the scaling behavior of the terms of the form 
$E R^N$. In our $D=5$ supergravity case, the contribution of these terms 
are similarly estimated to be 
\beqa
ER^N&\propto& \exp(2(N\!-\!1)\sigma), ~~~
\sigma=\sum_{a=0}^3 \beta^a.
\eeqa
By a change of basis using $U_{ia}$, the linear form $\sigma$ can be 
written in terms of simple roots of $G_2^H$ as
\beqa
\sigma=4\alpha_0 +3\alpha_1 +2\alpha_2 +1\alpha_3.
\eeqa
Therefore $(N-1)\sigma$ is always on the root lattice for any 
$N\in\mathbb{Z}$. Since our linear form $\sigma$ is invariant 
under permutation of spatial indices,  as is the case for the $D=11$ 
supergravity billiard, we expect that $(N-1)\sigma$ correspond to 
singlets in the $A_3$ decomposition for the correspondence between
$D=5$ supergravity and the $G_2^H$ sigma model to hold.
Amazingly, the series of singlets that we found in the last section 
have precisely 
such root coefficients $(\ell,m^1,m^2,m^3)=(4k,3k,2k,k)$! But to conclude 
that it can be interpreted as evidence for the conjectured correspondence, 
we must confirm if they have nonzero outer multiplicities. This is the task of 
the next subsection.

\subsection{Outer multiplicities}
For a given root of $G_2^H$, its outer multiplicity is computed 
as its root multiplicity 
minus the sum of weight multiplicities of representations 
which contains the root 
as a non-highest weight. For example, at $\ell=4$ there is a root 
$(\ell,m^1,m^2,m^3)=(4,3,2,1)$, which is an $A_3$ singlet 
$(p_1,p_2,p_3)=(0,0,0)$ 
with root multiplicity three. But at the same level there is another root 
$(\ell,m^1,m^2,m^3)=(4,4,3,2)$ with Dynkin label $(p_1,p_2,p_3)=(1,0,1)$, 
and, in fact, the representation with highest weight $(1,0,1)$ has a weight 
$(0,0,0)$ with weight multiplicity three, implying the outer multiplicity of 
$(4,3,2,1)$ is zero. 

On the other hand, at $\ell=8$ the root $(\ell,m^1,m^2,m^3)=(8,6,4,2)$ 
has root multiplicity 
21. Other (positive, $A_3$-dominant) roots at $\ell=8$ are
\begin{center}
\begin{tabular}{cccc}
$(\ell,m^1,m^2,m^3)$ &$(p_1,p_2,p_3)$& ~root multiplicity~&~weight contained
\\
$(8,6,4,2)$	     &$(0,0,0)$&$21$&\\
$(8,7,5,3)$	     &$(1,0,1)$&$9$&$3 (0,0,0)$\\
$(8,7,6,3)$	     &$(0,2,0)$&$3$&$(1,0,1) + 2 (0,0,0)$\\
$(8,8,6,3)$	     &$(2,1,0)$&$1$&$(0,2,0) + 2 (1,0,1) + 3 (0,0,0)$\\
$(8,7,6,4)$	     &$(0,1,2)$&$1$&$(0,2,0) + 2 (1,0,1) + 3 (0,0,0)$\\
\end{tabular}
\end{center} 
{}From the two rows at the bottom one can see that the outer multiplicity 
of $(0,2,0)$ is 
$3-1-1=1$. Then the outer multiplicity of $(1,0,1)$ is $9-2-2-1=4$. 
Finally the outer
multiplicity of the singlet is computed to be $21-3-3-2-4\times3=1$. 
Thus we see that 
there is a singlet at $\ell=8$. 

In principle one can similarly proceed to higher levels, but the computation 
becomes more tedious. Alternatively, one can use the equations of $(11.11)$ in 
\cite{Kac} to count the outer multiplicities directly. Namely, we expand the 
`denominator'  $F$ defined as\footnote{
In Exercise 11.11 of \cite{Kac}, $R$ should read $e^\rho R$, while 
it is correct in 1st and 2nd editions.
}
\beqa
R&=&\prod_{\alpha >0}(1-e^{-\alpha})^{{\rm mult}(\alpha)},\nonumber\\
F&=&-\log (e^\rho R),
\eeqa
not in the monomials of exponential of roots, but in the 
$A_3$ irreducible characters directly. To do this, we first write $F$ as
\beqa
F&=&-\rho+
\sum_{\alpha>0}{\rm mult}(\alpha)\sum_{n=1}^{\infty}
\frac{1}{n}e^{-n \alpha}\nonumber\\
&=&-\rho+
\sum_{n=1}^{\infty}(X|_{e^{-\alpha}\mapsto \frac{1}{n}e^{-n \alpha}}),
\eeqa
where 
\beqa
X&=&\sum_{\alpha>0}{\rm mult}(\alpha)e^{-\alpha}.
\eeqa
Since $X$ is invariant under the action of the $A_3$ Weyl subgroup, we may
expand $X$ in $GL(4)$ characters as 
\beqa
\label{Xins}
X&=&X_0+M_{(1,0,0,0)}\Theta_{(1,0,0,0)}
+M_{(0,1,0,0)}\Theta_{(0,1,0,0)}+\cdots,
\eeqa
where 
\beqa  
\Theta_{(i,j,k,n)}
=\genfrac{}{}{0.5pt}{}{\fr{D}_{(i+j+k+n,j+k+n,k+n,n)}}{\fr{D}_{(0,0,0,0)}}, 
\quad
\fr{D}_{(n_1,n_2,n_3,n_4)}=\det\big(x_i^{n_j+4-j}\big)
\label{GL4character}
\eeqa
is the $GL(4)$ character associated with the partition 
$(i+j+k+n+3,j+k+n+2,k+n+1,n)$, 
and
\beqa
X_0=e^{-\alpha_1} + e^{-\alpha_2} + e^{-\alpha_1 - \alpha_2} 
+ e^{-\alpha_3} + e^{-\alpha_2 - \alpha_3} + 
e^{-\alpha_1 - \alpha_2 - \alpha_3}
\eeqa
is the level $\ell=0$ piece; 
since it cannot be written as a character, we treat it 
separately. If we use the relation 
\beqa
(x_1,x_2,x_3,x_4)&=&
(e^{-\alpha_0},e^{-\alpha_0-\alpha_1},e^{-\alpha_0-\alpha_1-\alpha_2},
e^{-\alpha_0-\alpha_1-\alpha_2-\alpha_3})
\eeqa 
in (\ref{GL4character}),  then we find that 
$\Theta_{(p_1,p_2,p_3,p_4)}$ is precisely the $A_3$ character 
of Dynkin label $(p_1,p_2,p_3)$ 
at level $-\ell=-(p_1 +2p_2 +3p_3 +4p_4)$. Thus we can write $F$ as a linear 
function of  unknown coefficients $M_{(p_1,p_2,p_3,p_4)}$. 
Plugging this expression 
into the relation \cite{Kac}
\beqa
\label{Feq}
\sum_{i,j=0}^3 B_{ij}\Big(
\frac{\partial F}{\partial \alpha_i}\frac{\partial F}{\partial \alpha_j}-
\frac{\partial^2 F}{\partial \alpha_i \partial \alpha_j}\Big)=(\rho|\rho)
\eeqa
with
\beqa
B_{ij}&=&(\alpha_i|\alpha_j)~=~\left(\begin{array}{rrrr}
\frac23&-1&0&0\\
-1&2&-1&0\\
0&-1&2&-1\\
0&0&-1&2
\end{array}
\right)
\eeqa
and solving it by induction on the height,
we can compute outer multiplicities of roots directly. 
In this way we have computed outer multiplicities 
of all the $A_3$ representations 
appearing up to height $\leq 60$. Since it would not be very illuminating 
to list a number of pages of data on the decomposition, 
we will only display the list of 
outer multiplicities of the singlets in Table 3.

\renewcommand{\arraystretch}{1.2}

\begin{center}
Table 3. ~Outer multiplicities of the $A_3$ singlets. \\
\vskip 5mm
\begin{tabular}{|c|c|c|c|}

\hline
height&$(\ell,m^1,m^2,m^3)$&$(p_1,p_2,p_3)$&outer multiplicity
\\
\hline

10
&
$(4,3,2,1)$&$(0,0,0)$&
0 
\\
20
&
$(8,6,4,2)$&$(0,0,0)$&
1 
\\
30
&
$(12,9,6,3)$&$(0,0,0)$&
7 
\\
40
&
$(16,12,8,4)$&$(0,0,0)$&
59 
\\
50
&
$(20,15,10,5)$&$(0,0,0)$&
549 
\\
60
&
$(24,18,12,6)$&$(0,0,0)$&
5924 
\\
\hline
\end{tabular}

\end{center}


\section{Conclusions}
We have studied a one-dimensional sigma model 
associated with the hyperbolic Kac-Moody algebra $G_2^H$, and its 
possible link to $D=5$, ${\cal N}=2$ pure supergravity.
We have confirmed  the matching between the bosonic equations of motion 
and the Bianchi identity for $D=5$ pure supergravity and the equations 
of the $G_2^H$ sigma model with levels truncated to $\ell\leq 3$, 
to the same extent as the matching checked for the $D=11$ supergravity 
and the $E_{10}$ sigma model, in the sense that the terms that do not match 
have similar structures in both models and can be regarded as coming from 
roots with higher levels.

We have also studied the $A_3$ decomposition of $G_2^H$ at low levels. 
We have found three (presumably infinite) towers of roots which can be 
identified as the spacial gradients of the three lowest level fields. This is, 
again, the same observation as was already seen in the $A_9$ decomposition 
of $E_{10}$. We have also found a (again, presumably infinite) series of 
singlets at levels $\ell=4k$ for $k=2,3,\ldots$, 
which is consistent with the scaling 
of the higher curvature corrections of the form $ER^{k+1}$. 
This is a reasonable result because, 
in contrast to the M theory case, 
we expect corrections of the form $ER^{k+1}$  
for every positive integer $k$ in $D=5$ supergravity. 
Thus $D=5$ pure supergravity 
is the first example of theories with less supercharges 
than $D=11$ supergravity/M theory 
that shows evidence of the link between the higher order corrections 
and infinitely many roots of a certain Kac-Moody algebra. 

However, what is puzzling is the absence of any singlets at $\ell=4$, 
since this would 
predict the absence of $ER^2$ corrections in $D=5$ pure supergravity. 
The anomaly cancellation argument \cite{anomaly} 
analogous to the $D=11$ case 
requires that in $D=5$ pure supergravity there must be a gravitational 
Chern-Simons 
coupling proportional to $A\wedge \mbox{tr} R^2$ \cite{MO},
(Such a term is also known to arise in Calabi-Yau compactifications 
of M-theory \cite{FKM,FMS}.) 
and the super-invariant 
containing it will also have $ER^2$ terms. 
More explicitly, a general formula for 1-loop divergences in supergravity 
in diverse dimensions has been known for a long time \cite{FT}.  Using
this formula one can confirm that, unlike $D=11$ supergravity, 
the coefficient $\alpha_1$ of $R_{MNPQ}^2$ does {\it not} vanish for $D=5$ 
pure supergravity. Thus the correspondence 
between a singlet and a quantum correction fails at $\ell=4$, although there {\it are} 
three roots there.  
Despite this however,  the other higher level singlets which otherwise exist seem 
remarkable, and we may say that the hyperbolic Kac-Moody algebra $G_2^H$ has 
much chance of playing a crucial role in $D=5$ pure supergravity 
as $E_{10}$ has been conjectured to govern the dynamics of M theory.

Of course, 
$D=5$ supergravity is a non-renormalizable theory with little power 
of prediction 
for higher curvature corrections. 
If the conjecture is also true in $D=5$, what's the 
quantum counterpart of it? Although it is not so far known 
how to realize $D=5$ {\it pure} supergravity as a string compactification 
or any strong coupling limit thereof 
\cite{DH,MizoAsym}, 
turning on flux might cure this problem. More ambitiously, 
in view of the strong 
resemblance to the supergravity of distinguished character, 
it would be interesting 
to see if it can be realized as a Matrix-like theory \cite{BFSS,IKKT} 
with quarter supercharges.

\section*{Acknowledgments}
We are grateful to A. Kleinschmidt and H. Nicolai for valuable comments.
We also thank K-J. Hamada, K.~Ohashi and Y. Tanii for discussions.
S.M. thanks Max-Planck-Institute, Albert-Einstein-Institute for 
kind hospitality. S.M. and Y.Y. were supported by Grant-in-Aid
for Scientific Research (C)(2)\#16540273 and (B)\#17340047 from
The Ministry of Education, Culture, Sports, Science
and Technology.

%

\end{document}